\documentclass[12pt,preprint]{aastex}

\begin{document}

\title{Super-Keplerian Frequencies in Accretion Disks. Implications for
Mass and Spin Measurements of Compact Objects from X-ray Variability Studies}

\author{S.~A.~Mao}
\affil{Harvard-Smithsonian Center for Astrophysics, Cambridge, MA 02138; samao@cfa.harvard.edu}

\author{Dimitrios Psaltis\altaffilmark{1} and John A.\ Milsom}
\affil{Physics Department, University of Arizona, Tucson, AZ 85721}

\altaffiltext{1}{also, Astronomy Department, University of Arizona}

\begin{abstract} 
The detection of fast quasi-periodic variability from accreting black
holes and neutron stars has been used to constrain their masses,
radii, and spins. If the observed oscillations are linear modes in the
accretion disks, then bounds can be placed on the properties of the
central objects by assuming that these modes are locally
sub-Keplerian. If, on the other hand, the observed oscillations
correspond to non-linear resonances between disk modes, then the
properties of the central objects can be measured by assuming that
the resonant modes are excited at the same radial annulus in the
disk. In this paper, we use numerical simulations of vertically
integrated, axisymmetric hydrodynamic accretion disks to provide
examples of situations in which the assumptions implicit in both
methods are not satisfied.  We then discuss our results for the
robustness of the mass and spin measurements of compact objects from
variability studies.
\end{abstract}


\section{INTRODUCTION}

X-ray variability studies of accreting neutron stars and black holes
offer a tool for constraining the masses and spins of the compact
objects themselves. In the least model-dependent argument, the radial
extent of the region responsible for variability observed with a given
characteristic frequency is assumed to be at most equal to the size of
a Keplerian orbit with the same orbital frequency. Additionally, in
order for this region to be able to sustain long-lived oscillatory
modes, it must be larger than the radius of the innermost stable
circular orbit around the compact object. This requirement then places
an upper bound on the mass of the object and a lower bound on its spin
(see, e.g., Kluzniak, Michelson, \& Wagoner 1990; Mushotzky, Done, \&
Pounds 1993; Miller, Lamb, \& Psaltis 1998; Strohmayer 2001; for a
discussion see Psaltis 2005).

The assumption implicit in the above argument is that the Keplerian
frequency corresponds to the fastest dynamical timescale at any radius
in the accretion disk. This is usually justified by the fact that gravity
is the strongest force in the system (see, however, Alpar \& Psaltis
2008). Indeed, the study of normal modes in a viscous accretion disk
shows that the frequency of the lowest order, linear, hydrodynamic
mode excited at any radius has to be smaller than the local Keplerian
frequency (Kato 2001).

Observations of pairs of quasi-periodic oscillations from black holes
with frequencies in small-integer ratios, however, have challenged the
notion that these oscillations correspond to linear hydrodynamic modes
(Strohmayer 2001; Abramowicz \& Kluzniak 2001).  Such frequency ratios
are reminiscent of non-linear interactions, which can preferentially
amplify only those {\em high-order\/} harmonics of the normal modes
that are in resonance. If this is the prevailing mechanism for
compact-object variability, then meaningful bounds on or measurement
of their masses and spins can only be obtained within the context of a
detailed model for the oscillations. Indeed, assuming that the two
modes in resonance are dynamical hydrodynamic modes in the disk and using
an independent dynamical measurement of the mass of the black hole may
lead to an estimate of the black-hole spin (Abramowicz \& Kluzniak
2001; Abramowicz et al. 2003). The assumption implicit in this 
last argument is the fact that both modes are excited and trapped
in the same radial annulus in the accretion disk.

In this paper, we use numerical simulations of axisymmetric viscous
accretion disks in order to assess the validity of the two main
assumptions discussed above: that the highest frequency of oscillation
of a radial annulus in an accretion disk is the local Keplerian
frequency and that interactions may occur only between modes that are
locally excited. We show that neither of these two assumptions are
formally justified. 

We find that oscillatory modes that are excited close to the innermost
stable circular orbit generate traveling sound waves that propagate
radially to tens of Schwarzschild radii.  The presence of such modes
has been predicted in analytical studies of diskoseismic modes (Kato
2001). Our study shows, however, that they can grow to such amplitudes
that they can dominate the power spectrum of variability of the
accretion disks over extended regions. As viewed from any one radial
annulus outside their excitation region, these oscillations are highly
super-Keplerian. Moreover, non-linear couplings may lead to resonance
between the traveling modes and those that are excited locally. Both
results have important implications for the phenomenological
constraints imposed on the masses and spins of compact objects by the
observation of high-frequency quasi-periodic oscillations, which we
will discuss in detail below.

\section{ONE DIMENSIONAL MODELS OF HYDRODYNAMIC ACCRETION DISKS}

We used the numerical algorithm described in Milsom \& Taam (1996) in order
to solve the hydrodynamic equations that describe the time evolution of 
one-dimensional viscous accretion disks. For completeness, we summarize
here briefly the relevant equations and assumptions.

The conservation of mass is described by the continuity equation
\begin{equation}
\frac{\partial \Sigma}{\partial t}+\frac{\partial}{r\partial r}
(r\Sigma u_r)=0\;,
\end{equation}
where $\Sigma$ is the column density through the entire thickness of the
accretion disk at radius $r$ and $u_r$ is the radial velocity. The conservation
of linear and angular momentum are described, respectively, by the
equations
\begin{equation}
\frac{\partial (\Sigma u_r)}{\partial t}+
\frac{\partial}{r\partial r}(r\Sigma u_r^2)=
-\frac{\partial P}{\partial r}+\Sigma(\Omega^2-\Omega_{\rm K}^2)r
\end{equation}
and
\begin{equation}
\frac{\partial (\Sigma \Omega r^2)}{\partial t}+
\frac{\partial}{r\partial r}(r\Sigma u_r \Omega r^2)=
\frac{1}{r}\frac{\partial}{\partial r}\left(\Sigma \nu r^3
\frac{\partial \Omega}{\partial r}\right)\;,
\end{equation}
where $P$ is the sum of the vertically integrated gas and radiation 
pressures, $\Omega$ is the angular velocity in the flow, 
$\Omega_{\rm K}$ is the local azimuthal (Keplerian) dynamical 
frequency, and $\nu$ is the kinematic viscosity. Finally, the 
conservation of energy is described by the equation
\begin{equation}
\frac{\partial (\Sigma \epsilon_{\rm I})}{\partial t}+
   \frac{\partial}{r\partial r}\left[ru_r\left(\Sigma\epsilon_{\rm I}
+P\right)\right]=u_r\frac{\partial P}{\partial r}+\nu\Sigma
\left(r\frac{\partial \Omega}{\partial r}\right)^2
-\frac{4a c T^4}{3\kappa_0\Sigma}\;,
\end{equation}
where $\epsilon_{\rm I}$ is the internal energy per unit mass, $T$ is
the mid-plane temperature in the disk, $a$ is the radiation constant,
$c$ is the speed of light, and $\kappa_0$ is the sum of the
electron-scattering and free-free opacities of the flow.

This set of differential equations is augmented by a set of algebraic
equations that determine the gravitational potential of the central star,
the equation of state, and the kinematic viscosity in the flow.
We specify a pseudo-Newtonian gravitational potential through 
\begin{equation}
\Omega_{\rm K}=\sqrt{\frac{GM}{r(r-r_{\rm S})^2}}\;,
\end{equation}
where $r_{\rm S}\equiv 2GM/c^2$ is the Schwarzchild radius that 
corresponds to the mass of the star, $M$. We use an equation of state
that corresponds to a perfect gas in local thermodynamic equilibrium with
radiation such that
\begin{equation}
\epsilon_{\rm I}=\frac{3}{2}\frac{{\cal R}T}{\mu}
+\frac{2H a T^4}{\Sigma}
\end{equation}
and
\begin{equation}
P=P_{\rm gas}+P_{\rm rad}=\frac{\Sigma {\cal R}T}{\mu}+\frac{2}{3}H a T^4\;,
\end{equation}
where ${\cal R}$ is the gas constant, $\mu=0.62$ is the mean molecular
weight in the flow, $a$ is the radiation constant, and
\begin{equation}
H\equiv\frac{\sqrt{P/\Sigma}}{\Omega_{\rm K}}
\end{equation}
is the vertical scale height of a disk in hydrostatic equilibrium.
Finally, we use the $\alpha$-prescription 
\begin{equation}
\nu=\frac{2}{3}\alpha\frac{P_{\rm gas}}{P}\sqrt{\frac{P}{\Sigma}}H
\end{equation}
for the kinematic viscosity, which ensures thermal and viscous stability.

In all numerical simulations we set the mass of the central object to
10~$M_\odot$ and specified the solution by the mass accretion rate
$\dot{M}$ in units of the Eddington accretion rate $\dot{M}_{\rm
E}\equiv 64\pi GM/(k_{\rm es}c)$ and the viscosity parameter $\alpha$.
The domain of solution, initial conditions, boundary conditions, and
method of solution are described in detail in Milsom \& Taam (1996).

The global timing properties of our numerical simulations have been
described in detail by Milsom \& Taam (1996). In summary, for a given
value of the viscosity parameter $\alpha$, simulations with low
(typically $\lesssim 0.01 \dot{M}_{\rm E}$) accretion rates result in
lightcurves with noisy power spectra, simulations with intermediate
accretion rates exhibit large amplitude global modes at frequencies
comparable to the maximum radial epicyclic frequency in the flow and
its harmonics, and simulations at even larger accretion rates
(typically $\gtrsim 0.02 \dot{M}_{\rm E}$) exhibit no significant
variability (although epicyclic oscillations were always visible in
subsequent 2D hydrodynamic simulations; see Milsom
\& Taam 1997).

\begin{figure}[t]
 \centerline{
\plotone{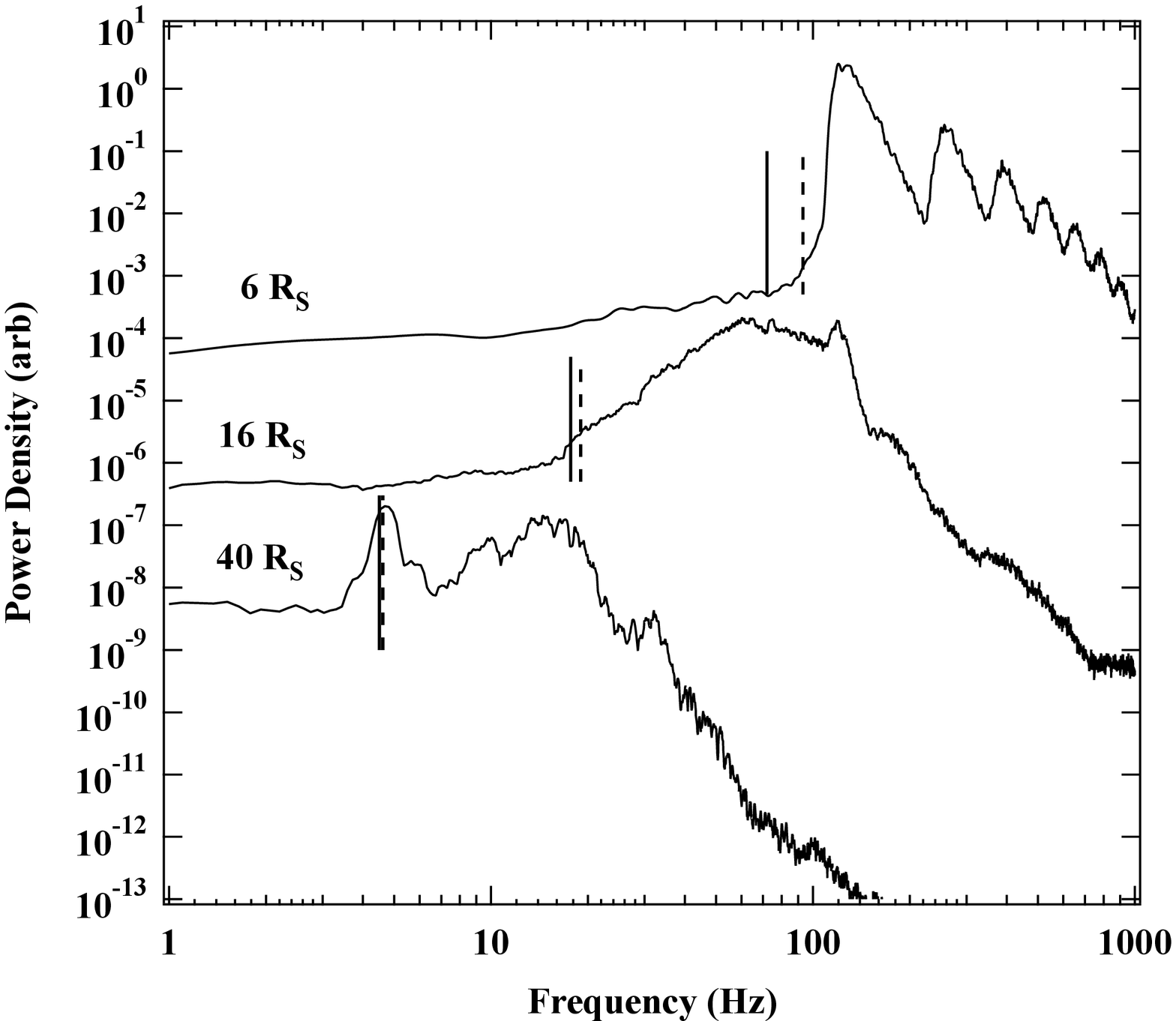}}
  \caption{\footnotesize
The power spectrum of the vertically integrated column density $\Sigma$ at
three different radii in an accretion disk with an accretion rate of
$0.1\dot{M}_{\rm E}$ and a viscosity parameter $\alpha=0.1$. The power
spectra have been displaced vertically for clarity. The vertical solid
line on each power spectrum corresponds to the local radial epicyclic
frequency, whereas the vertical dashed line on each power spectrum
corresponds to the local Keplerian frequency.} 
\label{fig:pds}
\end{figure}

\section{SUPER-KEPLERIAN FREQUENCIES IN THE NUMERICAL SIMULATIONS}

The main goal of our study is to investigate the radial propagation of
the traveling modes excited in the various simulations, in order to
quantify the extent of the domain in which they dominate the
variability. For this reason, we performed Fast Fourier Transforms
(FFTs) of a number of physical quantities such as the radiation flux,
local mass accretion rate, vertically integrated column density, etc.,
evaluated at different radii in the flow. In order to resolve the
fastest possible oscillations we used a time resolution of 2~ms, and
in order to resolve the power-spectral peaks of potential oscillations
of high coherence we used segments of 4~s length. Finally, following
standard procedure, we averaged the FFTs of four sequential segments
in every simulation, in order to improve the statistics, and binned
the power density spectra in frequency space. Hereafter, we will be
discussing the power density spectra of the vertically integrated
column density in the disk.

Figure~\ref{fig:pds} shows a typical example of the power spectrum of
the vertically integrated density $\Sigma$ evaluated at three
different radii in the accretion disk for the simulation with an
accretion rate of $0.1 \dot{M}_{\rm E}$ and a viscosity parameter of
$\alpha=0.1$. We can clearly see that the accretion disk is variable
over a wide range of timescales and, similar to the observed systems,
exhibits both broad-band noise and quasi-periodic
oscillations. Indeed, at $6R_{\rm S}$ a large number of oscillation
peaks can be seen, most of which are overtones to a fundamental
frequency.  However, contrary to all expectations, most of the
quasi-periodic oscillations that appear in the power spectra occur at
frequencies larger than the local Keplerian frequencies, indicated in
the Figure with the vertical dashed lines.

\begin{figure}[t]
 \centerline{
\plotone{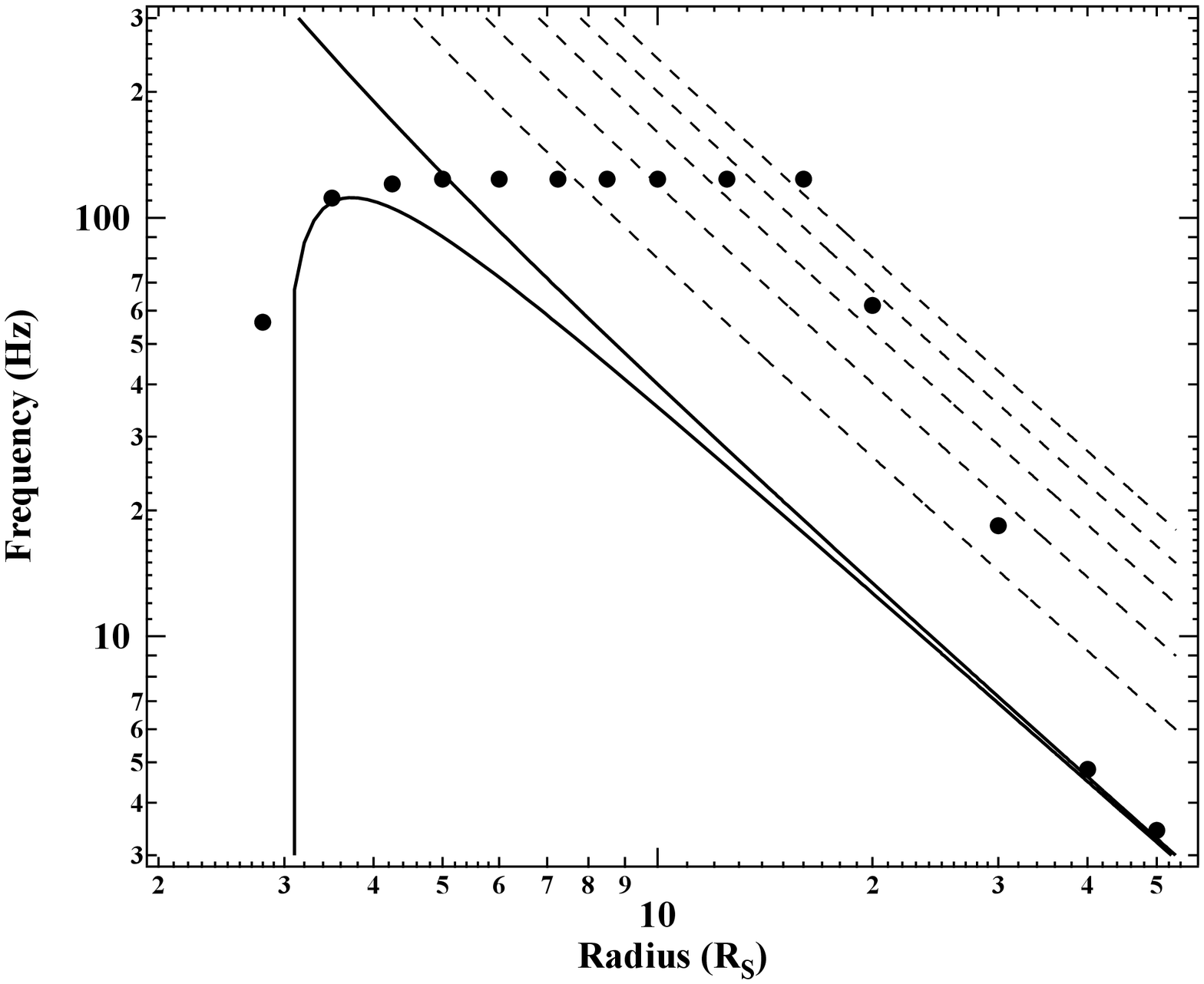}}
  \caption{\footnotesize
The radius dependence of the dominant oscillation frequency
of the vertically integrated density in a simulation with 
$\dot{M}=0.1\dot{M}_{\rm E}$ and a viscosity parameter $\alpha=0.1$.
At each radius, the solid lines show the radial epicyclic and Keplerian
frequencies, whereas the dashed lines show increasingly higher
harmonics of the local Keplerian frequency.}
\label{fig:freq}
\end{figure}

Figure~\ref{fig:freq} shows the frequency of the dominant
quasi-periodic oscillation of the vertically integrated density as a
function of radius, for the simulation with $\dot{M}=0.1\dot{M}_{\rm E}$ and
a viscosity parameter $\alpha=0.1$. The radius dependence of the
oscillation frequency depicted in this figure shows three distinct
regions that appear in varying degrees in all simulations. First,
there is a wide region in the inner accretion disk where the dominant
oscillation frequency is constant and nearly equal to the maximum of
the radial epicyclic frequency. This is surrounded by a second region
in which the oscillation frequency is radius-dependent, rapidly
decreasing, but super-Keplerian. Finally, in a third, outer region,
the oscillation frequency is radius-dependent and nearly equal to the
radial epicyclic (or local Keplerian) frequency.

\begin{figure}[t]
 \centerline{
\plotone{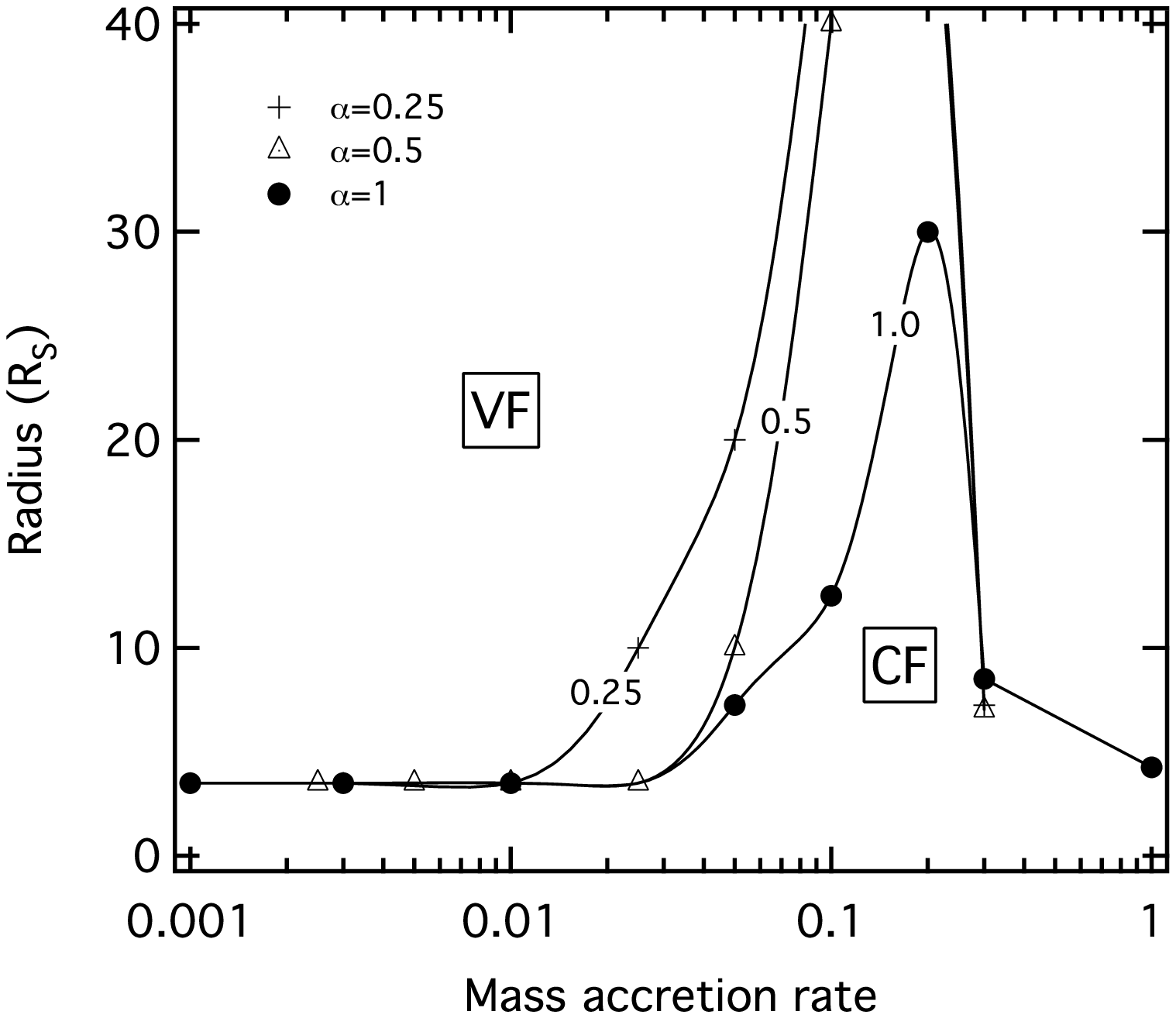}}
  \caption{\footnotesize
  The regions in the accretion disk where the
  constant-frequency (CF) and the variable-frequency (VF) modes live 
  for simulations with different mass accretion rates and viscosity 
  parameters $\alpha$.}
 \label{fig:rad}
\end{figure}

The radial extent of the regions in the accretion disk in which the
constant-frequency and variable-frequency modes live depends strongly
on the mass accretion rate and rather weakly on the viscosity
parameter $\alpha$, as shown in Figure~\ref{fig:rad}. At accretion
rates smaller than a few hundredths of the Eddington critical value,
the oscillation frequencies at all radii are nearly equal to the local
epicyclic frequencies. As the accretion rate increases beyond $\sim
0.01 \dot{M}_{\rm E}$, the radial extent of the constant-frequency mode also
increases, reaching radii $\gtrsim 40 R_{\rm g}$ for accretion rates
comparable to $0.1 \dot{M}_{\rm E}$; beyond that radius, only the
variable frequency mode exists.

%\begin{figure}[t]
% \centerline{
%\psfig{file=frequency.ps,angle=0,width=10truecm}}
%  \caption{\footnotesize
%The ratio of the frequency of the dominant oscillation
%at a radius of 25$R_{\rm g}$ to the local radial epicyclic frequency
%$\kappa$, for simulations with different values of the mass accretion 
%rate and the viscosity parameter $\alpha$.}
%\label{fig:freq_ratio}
%\end{figure}

%The frequencies of the variable-frequency oscillations are dictated by
%the radial extent of the region in which the constant-frequency mode
%lives. This is illustrated in Figure~(\ref{fig:freq_ratio}), which
%shows the frequency of the dominant mode at a radius of $25R_{\rm g}$
%for simulations with different values of the mass accretion rate and
%the viscosity parameter. 

\begin{figure}[t]
 \centerline{
\plotone{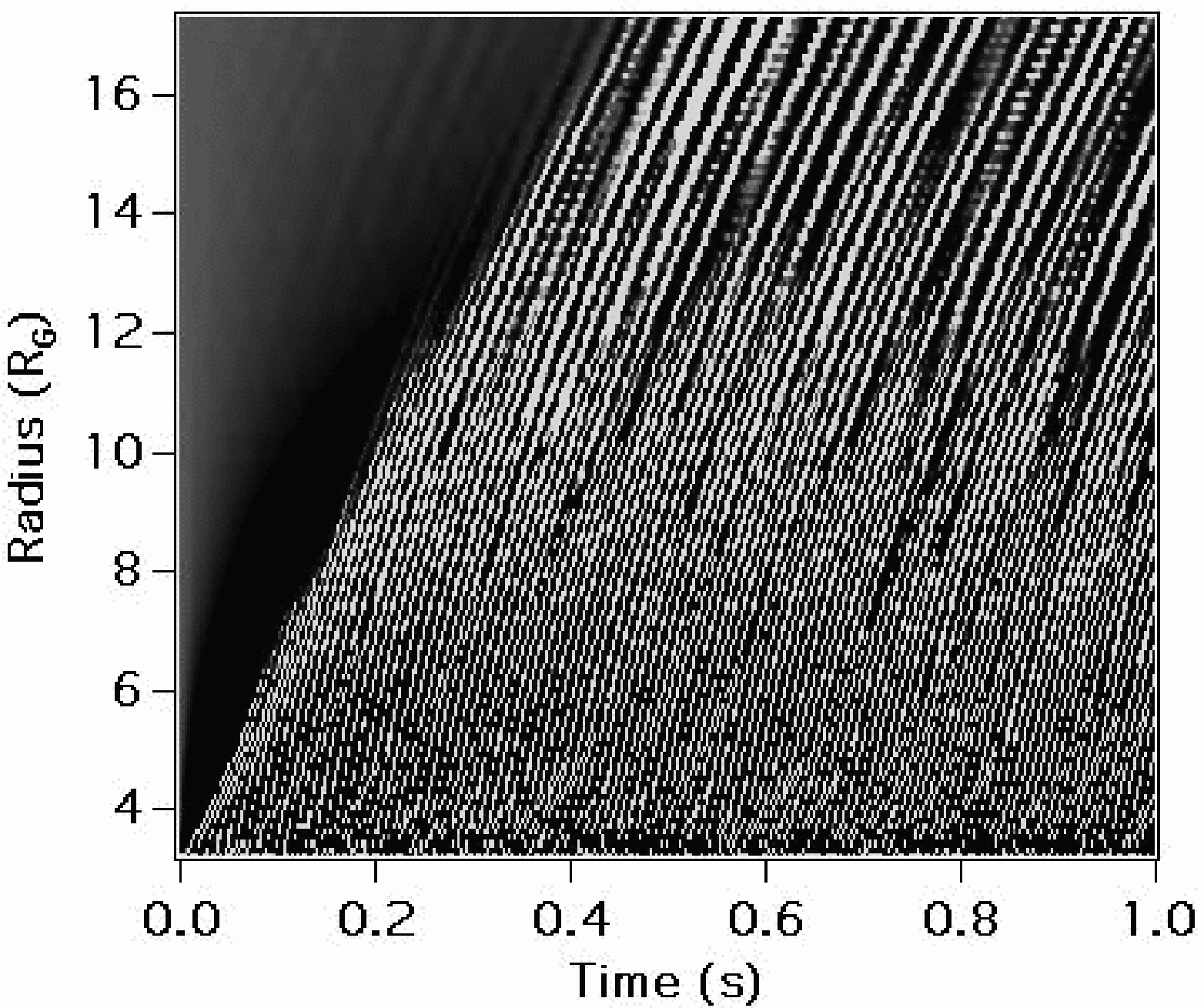}}
  \caption{\footnotesize
A radius-time plot of the vertically integrated density for
the simulation with an accretion rate of 0.1$\dot{M}_{\rm E}$ and a 
viscosity parameter $\alpha=0.1$.}
\label{fig:radiustime}
\end{figure}

The constant frequency modes that dominate the variability in the
inner accretion disk at accretion rates $\simeq 0.1 \dot{M}_{\rm E}$
are neither excited nor trapped in that region. They are in fact
traveling inertial-acoustic modes that are generated close to the
innermost stable circular orbit and propagate outwards (see also Chen
\& Taam 1995). This is illustrated in Figure~\ref{fig:radiustime},
which shows the radius-time plot of the vertically integrated column density
for the simulation with an accretion rate of 0.1$\dot{M}_{\rm E}$ and
a viscosity parameter $\alpha=0.1$. 

\section{DISCUSSION}

We investigated the presence of oscillatory radial modes in numerical
simulations of axisymmetric, viscous accretion disks. In agreement
with earlier efforts (see Chen \& Taam 1995), we find that traveling
modes are generated close to the innermost stable circular orbit of
the flow and propagate outwards to several tens of Schwarzschild
radii, without getting attenuated. As viewed from any given annulus in
the accretion disk, these waves appear as density oscillations with
frequencies that are locally super-Keplerian.

The presence of such super-Keplerian oscillations in the simulations
implies that we cannot use the frequency of an observed oscillation to
set a bound on the extent of the region in the accretion flow that is
responsible for the modulated emission. For example, if we were to
assume, as is normally done, that the oscillation with a frequency
$f\simeq 110$~Hz shown in Figure~\ref{fig:freq} is limited by the
local Keplerian frequency, i.e., that $f\le f_{\rm K}\simeq
(1/2\pi)\sqrt{GM/R^3}$, then we would infer that the region responsible
for this oscillation is smaller than
\begin{equation}
R\le 4.7 R_{\rm S} \left(\frac{M}{10~M_\odot}\right)^{1/3}
\left(\frac{f}{110~{\rm Hz}}\right)^{-2/3}\;.
\end{equation}
However, this limit would seriously underestimate the radial extent
of the region where the $\simeq 110$~Hz mode resides, which is in fact
four times larger (see Fig.~\ref{fig:freq}). This is a direct consequence
of the fact that the traveling mode is locally super-Keplerian in the
entire inner region of the disk.

It is important to emphasize here that, although the
constant-frequency modes are super-Keplerian in the region of the
accretion disk where they are traveling, they are sub-Keplerian near
the innermost stable circular orbit, where they are excited. This is
true for all modes in all simulations that we explored.  As a result,
even though we cannot use the frequency of the mode to set an upper
bound on the radial extent of the region in which it resides, we can
still constrain the mass and the spin of the compact object by
requiring that the mode is sub-Keplerian in the region where it is
excited.

Our simulations can only recover the excitation and propagation of
axisymmetric radial oscillations. Because of this, we cannot simulate
the possible non-linear coupling of these to other, azimuthal or
vertical, modes. However, the presence of large amplitude traveling
modes in the simulated disks suggests that the non-linear coupling and
resonances may not necessarily occur between modes that are excited in
the same region in the accretion disk. Indeed, Figure~\ref{fig:freq}
shows that the $\simeq 110$~Hz mode is in 2:1 resonance with an
azimuthal Keplerian mode at $\simeq 7 R_{\rm S}$, in 3:2 resonance
with an azimuthal mode at $\simeq 8.5 R_{\rm S}$, etc. As a result, it
is not justifiable to use the frequencies of observed resonant
oscillations and infer the properties of the compact objects assuming
that they correspond to modes in the same region in the accretion flow
(as in, e.g., Abramowicz \& Kluzniak 2001).

Finally, it is worth pointing out that our results are valid for
geometrically thin, hydrodynamic accretion disks with an alpha
viscosity.  Traveling inertial-acoustic modes such as those we studied
here are not visible in time-dependent simulations of
magnetohydrodynamic accretion disks (see, e.g., Hawley \& Krolik 2001;
Armitage \& Reynolds 2003; but see also Arras, Blaes, \& Turner 2006
and Chan et al.\ 2008). This is probably due to the combined effects
of the magnetorotational instability, which damps such radial
epicyclic modes, and the presence of fully developed turbulence that
quickly destroys the coherence of such traveling modes. It is not
clear, at this point, what physics needs to be included (or perhaps
what assumptions need to be relaxed) for the numerical simulations of
magnetohydrodynamic disks to show large-amplitude oscillations that
are ubiquitous in nature. Independent of whether the hydrodynamic
inertial-acoustic modes survive in realistic simulations, our results
still serve as a proof of principle that high amplitude,
super-Keplerian traveling modes may, in principle, exist in the inner
regions of accretion flows.

\acknowledgements

D.\,P.\ was supported in part by the NSF CAREER award NSF 0746549.

\end{document}